\documentclass[aps,prb,twocolumn,preprintnumbers,amsmath,amssymb,reprint,longbibliography]{revtex4-1}
\usepackage{graphicx}
\usepackage{dcolumn}
\usepackage{bm}
\usepackage{amsmath}
\usepackage{feynmp}

\usepackage{graphicx}

\begin{document}

\title{Modification and Control of Topological Insulator Surface States Using Surface Disorder}

\author{Vincent Sacksteder}
\email{vincent@sacksteder.com}
\affiliation{School of Physical and Mathematical Sciences, Nanyang Technological University, 637371, Singapore}

\author{Tomi Ohtsuki}
\affiliation{Department of Physics, Sophia University, Tokyo, 102-8554, Japan}

\author{Koji Kobayashi}
\affiliation{Department of Physics, Sophia University, Tokyo, 102-8554, Japan}


 




\begin{abstract}
We numerically demonstrate a practical   means of systematically  controlling topological transport  on the surface of a three dimensional topological insulator, by introducing  strong disorder in a layer of depth $d$ extending inward from the surface of the topological insulator.   The dependence on  $d$ of the density of states, conductance, scattering time, scattering length,  diffusion constant, and mean Fermi velocity are investigated. 
The proposed control via disorder depth $d$ requires that the disorder strength be near the large value  which is necessary to drive the TI into the non-topological phase.   If $d$ is patterned using masks, gates, ion implantation, etc., then integrated circuits may be fabricated.  This technique will be useful for experiments and for device engineering.
\end{abstract}

\maketitle



\section{Introduction}

Recently a new kind of material has been predicted and measured: topological insulators, which do not permit current to flow through their interior but do allow  metallic conduction along their surfaces.  \cite{Kane05, Zhang09, Hasan10, Li12, Culcer12}  The conducting states residing on the topological insulator (TI) surface
are protected topologically, meaning that they are safeguarded by the bulk's insulating property from local perturbations as long as a mobility gap is maintained between the bulk bands.  In consequence the surface states are only weakly sensitive to fine details of the bulk Hamiltonian, such as lattice structure, details of atomic bonding, or interactions.  They are, however, vulnerable to any conduction through the bulk at energies lying in the bulk band gap, and  therefore  demand a very high-purity bulk. Any engineering of TI devices  for either practical or scientific applications  will likely use a very pure and unaltered  TI bulk, and utilize only  the TI's surface.

There are several compelling motivations for using TI surfaces instead of conventional materials to carry current.  Power dissipation may  be reduced by the TI's robustness against disorder.   TIs also lock spin to momentum, which will be useful  for creating spin-polarized currents and for conducting spin over long distances.  They may also host exotic states that could be used for quantum computing, such as  Majorana fermions or strongly interacting topological phases.  \cite{fu2008superconducting, PhysRevB.83.195139,  PhysRevLett.105.246809} However these applications all suffer from the topological state's resilience against the mechanisms usually employed to direct or switch off electronic conduction, such as gating and etching.   TI surface states are  difficult to control.

In this article we  propose a way of  engineering TI device properties
 to match  engineering requirements.   
Our main contribution  is the observation  that  introducing disorder \textit{only near the TI's surface},  in a region beginning on the TI surface and extending inward  to a depth  $d$, 
is a practical way of controlling the surface states.   
As the electron moves across the disordered surface of the TI,  from time to time it becomes almost trapped at a particular site and dwells there for a while before continuing its journey.  This trapping is unable to destroy or localize the in-gap surface state, but it does cause  a localized increase in the surface state's probability density, and our numerical results show a corresponding increase in the in-gap density of states $\rho$.  Moreover, the increased dwell times at individual sites cause, on average, a decrease in the Fermi velocity $v_F$.    The altered density of states and Fermi velocity change also  the  scattering time $\tau$, diffusion constant $D = v_F^2 \tau/2$, and coupling constant $\alpha = e^2/ \epsilon \hbar v_F$ controlling interactions. 

The impact of surface disorder on a topologically protected state  is ordinarily limited by the state's  tendency to shift into the clean bulk to avoid disorder.    However we can trap the topological state in the disordered region   by tuning  the disorder strength $W$  near the critical value $W_c$ which causes  the disordered TI to transit from the topological phase to the non-topological phase.

Our main point is to  demonstrate that when the topological state is trapped in the disordered region it is strongly sensitive to  the disorder depth $d$.  
This is a unique control parameter which, if patterned on a TI surface,  can create  areas with slower conduction and increased  density of states,  introduce control points that are sensitive to  gating voltages,
and direct conduction along particular channels.

We emphasize that there are several practical and achievable techniques for producing patterned surface disorder.  Already many experiments have studied the effects on TI surfaces of disorder induced by atmosphere, by deliberate introduction of adsorbed molecules and dopants, and even by mechanical surface abrasion. \cite{Hsieh09, bianchi2010coexistence, Analytis10, Noh11, Kong11, Tereshchenko11, Brahlek11, Liu12, Aguilar12, kim2013surface}  Numerous experiments have also demonstrated that capping can effectively protect a TI surface, so masking and etching techniques are promising, as is ion implantation.   \cite{Lang11} 

Ion implantation in particular gives  precise control of impurity concentration and depth, and allows  control of the Fermi level by mixing ions.  This technique has been been developed extensively for applications to semiconductors, but its application to TIs is at an exploratory stage.  The related technique of ion milling, useful for controlling sample thickness and for polishing the surface, was first applied to TIs in 2010 and is now in widespread use. \cite{li2010van}  Ion implantation was first applied to TIs  in studies of the spin response to a magnetic field \cite{vazifeh2012spin,PhysRevB.90.214422}, and more recently has been used to dope the Fermi level \cite{sharma2014ion} and to add disorder to the TI surface. \cite{PhysRevB.91.085107}

\begin{figure}[tbp]
\begin{tabular}{c}
\includegraphics[width=8.5cm,clip,angle=0]{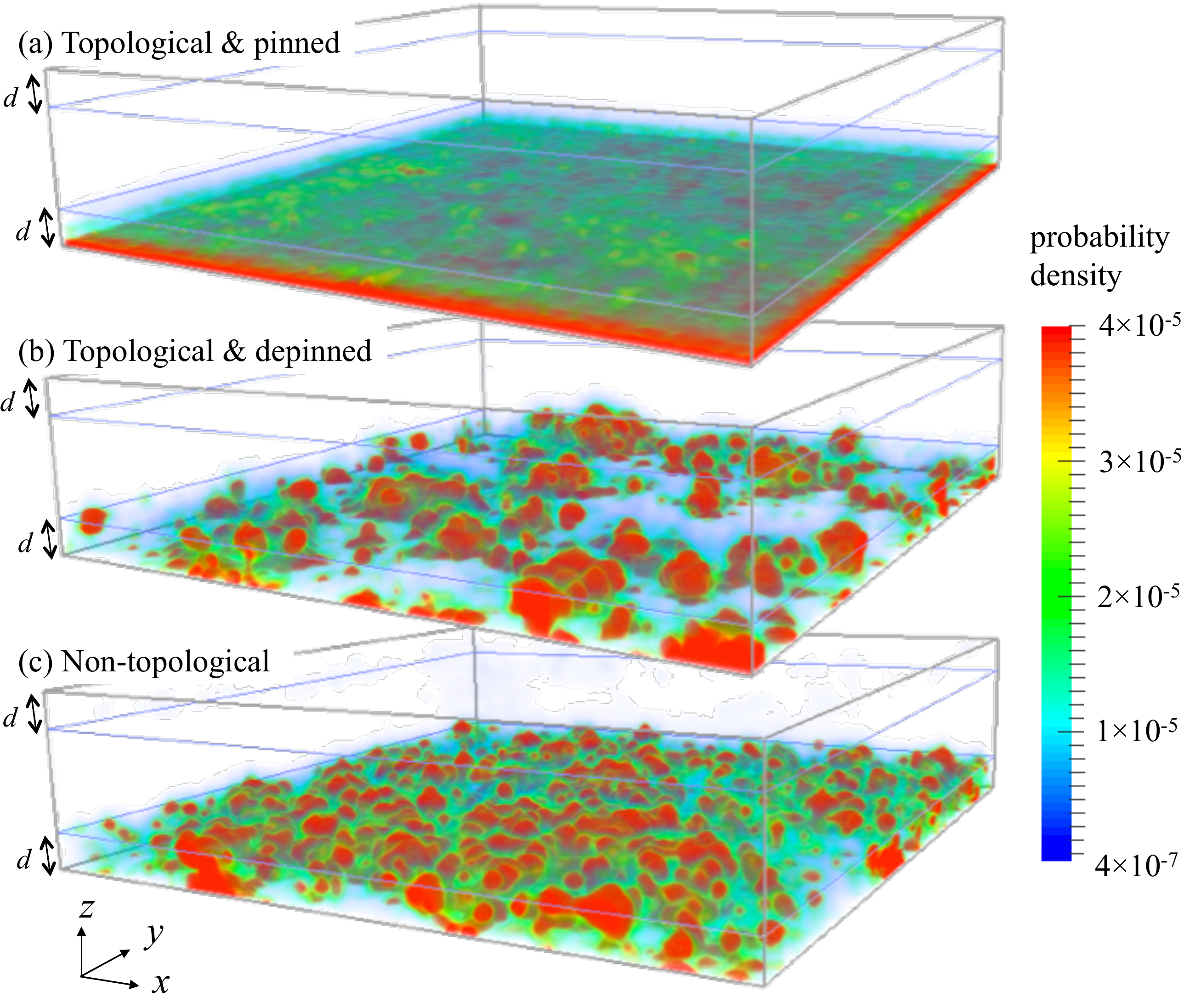}
\end{tabular}
\vspace{-2mm}
\caption{The three types of transport illustrated in a disordered surface layer of depth $d=5$.  Panel (a) shows the probability density $ | \psi |^2$ at small disorder $W=4.5$; the topological state is pinned to the sample's bottom surface.  Panel (b) shows a topological state at $W=7.5$,  near $W_c$.  It is fully depinned from the bottom surface but nonetheless supports topologically protected conduction.   Panel (c) shows strong disorder $W = 9.0$, where the disordered region is no longer topological and hosts the  bulk states seen here. A topological state at the disorder boundary coexists with these bulk states, and both contribute to the conductivity. The sample dimensions are $l \times w \times h=80 \times 80 \times 20$, with periodic boundary conditions in the $x$ and $y$ directions, and  $E_F \approx 0$.  
}
\label{Eigenfunctions}
\end{figure}

The surface state control which is proposed here  is obtained only at the critical disorder $W \approx W_c$, which is generally quite large, roughly the same as the bulk band width $E_B$, but can be reduced to much smaller values by tuning the Fermi level.
  \cite{Groth09,Xu12} 
At small disorder $W \ll W_c$ the topological state is pinned at the outer boundary of the disordered region, as seen in Figure \ref{Eigenfunctions}a. Its conduction is  therefore insensitive to the region's depth.   In the opposite case of  large disorder  $W \gg W_c$ the disordered region fills with non- topological states deriving from the bulk band (Figure \ref{Eigenfunctions}c) \cite{Ringel12, Schubert12, wu2013robust, wu14}   which contribute to conduction, and  the conductance and the density of states both depend on $d$.  
 Only near $W \approx W_c$  does the topological state depin from the TI's outer surface and stretch inward to the interface with the clean bulk (Figure \ref{Eigenfunctions}b), producing conduction which is both  topological and sensitive to the disorder depth $d$.

In Section \ref{StateAnalysis} we analyze  these three types of transport  and determine in each case how the transport  parameters scale with the disorder depth $d$.  Next section \ref{NumericalResults} presents our numerical model and confirms numerically the existence of the depinned topological state  whose transport  can be controlled by disorder depth.  We conclude with section \ref{Applications}, which discusses how to pattern topological conduction to meet device requirements.  

\section{Scaling Analysis of Disordered Transport\label{StateAnalysis}}  Depending on the disorder strength, three different types of conducting states  may be obtained.  These states have clear signatures in the dependence of the conductance $G$ and the 2-D density of states $\rho_{2D}$ on disorder depth $d$, which we show in Table \ref{table:scaling} and will exhibit in our numerical results.  The \textit{pinned topological  state} at $W \ll W_c$ is not sensitive to $d$.  At the other extreme $W \gg W_c$  \textit{non-topological states}  in the disordered region are important.    The scattering length $l$ of these strongly disordered states is comparable to the lattice spacing and much smaller than the disorder depth, so these states undergo true 3-D diffusive transport, with $G$ and $\rho_{2D}$ both linear in $d$. 

In contrast, the \textit{depinned  topological state} at $W \approx W_c$
can be identified clearly by  the signature of a $d$-linear 2-D DOS $\rho_{2D}$, combined with a constant conductance $G$.  The linear $\rho_{2D}$ is   caused by the state depinning from the TI surface, while the constant $G$ controverts nontopological conduction, which would exhibit 3-D diffusion and a linear conductance.   

This signature is unambiguous.  The linear DOS cannot be attributed to any non-topological surface or bulk states distinct from the depinned topological state, because all in-gap states are located in the disordered region, which near $W_c$ delocalizes and hosts only  extended states like that seen  in Figure \ref{Eigenfunctions}b.   
 In the parallel case of 2-D TIs with edge disorder
 very few in-gap states are localized, and the conductance remains quantized. \cite{wu2013robust, Stanescu09,Buchhold12}  In both 2-D and 3-D at $W \approx W_c$ all states mix and participate in the topological conduction. 


  The depinned state is very remarkable for being simultaneously robustly conducting and extremely disordered.  This is seen clearly in Figure \ref{Eigenfunctions}, where we have calculated the states' inverse participation ratios (IPR), a measure of their volumes defined by $\sum_i(\sum_{\alpha=1}^4 |\psi(i,\alpha)|^2 )^2$ with
  $\psi(i,\alpha)$ the $\alpha$-th component of the wave function on the site $i$.  These eigenfunctions were obtained by diagonalizing the Hamiltonian using the sparse matrix diagonalization subroutine FEAST built with the Intel Fortran Math Kernel Library.  The system size is $80 \times 80 \times 20$, and periodic boundary conditions have been imposed in $x$ and $y$ direction, while the open boundary condition is imposed in $z$ direction.
  The IPR of the wave function of depinned state (Fig. 1(b)) is  $2.09\times 10^{-3}$,
an order of magnitude larger than those of the pinned and bulk states (Fig. 1 (a) and (c)),
which are $1.25\times 10^{-4}$ and $1.73\times 10^{-4}$, respectively.  
  In non-topological systems this type of volume reduction would be accompanied by Anderson localization, i.e. the state's extent along the $x$ and $y$ axis parallel to the surface would be much smaller than the sample size.  Here topology ensures that the depinned state  remains conducting and extended across the entire sample.

 \begin{table}[htb]
\begin{center}
\caption{The transport parameters' scaling with disorder depth $d$. $G$ is the conductance, $\rho_{2D}$ is the density of states, $v_F$ is the mean Fermi velocity, $\tau$ is the scattering time, $l$ is the scattering length, $D$ is the diffusion constant, and $\alpha$ is the dimensionless coupling constant governing interactions.}
\begin{tabular}{lllllllllll}
  \hline \hline
  Conducting State & $\; G \;$ & $\rho_{2D}\;$ & $v_F\;$ \; & $\tau \;$  &  $l = v_F\tau\;$ & $ D\;$ & $\alpha\; $\\
   \hline
   topological \& pinned  & $1$ & $1$ & $1$ & $1$ &  $1$ & $1$ & $1$ \\
 topological \& depinned & $1$ & $d$ & $1/d$ &  $d$ &  $1$ & $1/d$ & $d$ \\
non-topological & $d$ & $d$ & $1$ & $1$ & $1$ & $1$  & $1$ \\
  \hline \hline
\end{tabular}
\label{table:scaling}
\end{center}
\end{table}

The depinned state is topologically guaranteed to conduct over large distances,  and therefore must have a long-wavelength limit where the average parameters of  2-D surface transport  are well-defined, including the 2-D density of states $\rho_{2D}$, average Fermi velocity $v_F$, etc. \cite{wu2013robust}  We will show that these quantities are strongly dependent on $d$, beginning  with the Fermi velocity $v_F = dE/dk$, the eigenvalue's derivative with respect to $k$.  Its scaling can be determined from the fact that the depinned topological state is not localized.  Therefore the eigenvalues within the gap repel each other according to Wigner-Dyson level repulsion, and the energy scale $dE$ in $v_F = dE/dk$ is set by the level spacing $\Delta E$.  \cite{wu2013robust} The depinned state extends inward to depth $d$, so its 2-D DOS   $\rho_{2D}$ is proportional to $d$, and   $\Delta E$ scales with $1/d$. Since in strongly disordered samples the momentum scale $dk$  in $v_F$ is controlled by the inverse of the lattice spacing $a$ and is not sensitive to $d$, we obtain $v_F \propto 1/d$.

Next we note that at $W \approx W_c$ the scattering length $l$ is close to the lattice spacing and independent of $d$.  Since $l = v_F \tau$, the scattering time grows linearly with $d$, inversely to $\Delta E$.    We report these scaling relations, along with  $D = v_F^2 \tau /2 $ and the dimensionless coupling constant  $\alpha = \frac{e^2}{\epsilon \hbar v_F}$ which controls interaction effects, in Table \ref{table:scaling}.  \cite{neto2009electronic}  Table \ref{table:scaling}'s results for the depinned state can be determined  directly from dimensional analysis by finding each quantity's scaling with either the scattering time $\tau \propto d$ or its inverse, the energy $\propto 1/d$.   In summary, $v_F, \tau, D,$ and $\alpha$ are all very sensitive to the disorder depth $d$, while $G$ and $l$ are not.


\begin{figure}[tbp]
\begin{tabular}{c}
\includegraphics[width=7cm,clip,angle=0]{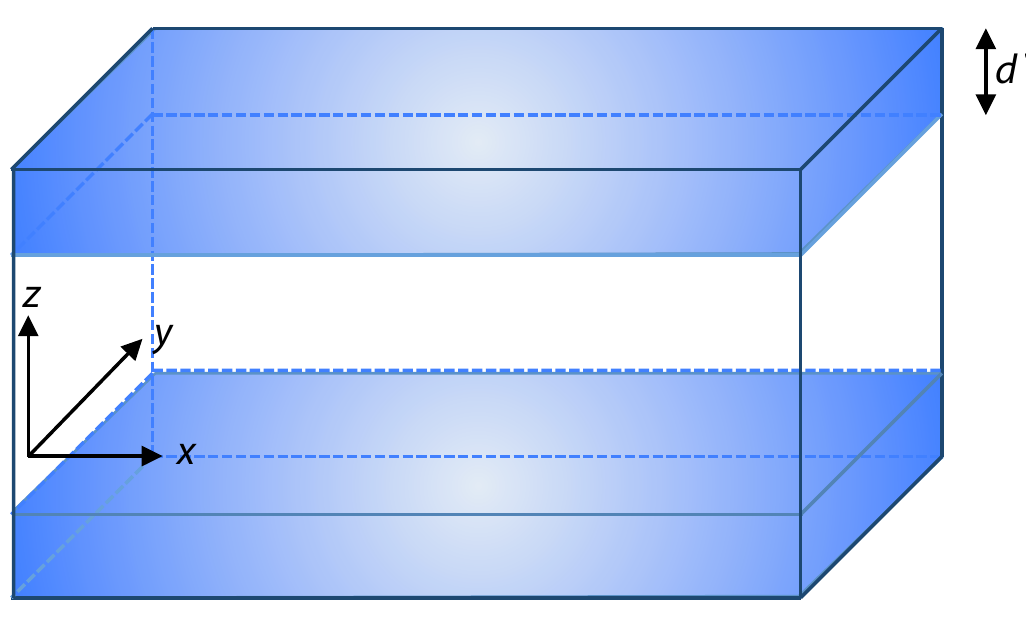}
\end{tabular}
\vspace{-2mm}
\caption{Schematic of the sample geometry.  The shaded region is disordered.  In the
calculation of the density of states, periodic boundary conditions (pbc)
are imposed in the $x$- and $y$-directions.  In the  conductance
calculation, pbc are imposed in the $y$-direction and current flows in the $x$ 
direction.  In both cases, fixed boundary conditions are imposed in the
$z$-direction.
}
\label{Schematic}
\end{figure}

\section{Numerical Results\label{NumericalResults}} We turn to  calculations of the effect of disorder depth $d$ on $\rho$ and $G$ which will  confirm numerically the existence of a depinned topological state with the scaling listed in Table \ref{table:scaling}. The topological surface conduction which interests us is  independent of 
 any microscopic details of the Hamiltonian.  
 Therefore we study a  computationally efficient  minimal tight binding model of a strong  $\mathbb{Z}_2$  TI implemented on a cubic lattice.
 We leave the TI bulk pure, since the main effects of  bulk disorder    can be duplicated  by narrowing the bulk band gap and increasing the penetration depth in a pure TI.\cite{Groth09,Guo10,Chen12,Xu12,kobayashi2013disordered}   We also omit the effects of bulk defects, which are known to dope the Fermi level toward the conduction or valence band depending on the defect type, and at sufficient concentrations also cause a conducting impurity band to be formed inside the bulk band gap. \cite{lee2014stability} We will return to doping in section \ref{Applications}. 
  With four orbitals per site, the model's momentum representation is:
   \begin{align} \label{eqn:H}
      H = &  \sum_{i=1}^3 \left[\left(\imath  \frac{t}{2} \alpha_{i} 
                                         -\frac{1}{2}   \beta\right)e^{-\imath k_i a} + \rm{H.c.}\right]   + (m+3 ) \beta  
   \end{align}
\noindent $\alpha_{i} = \sigma_x \otimes \sigma_i$ and $\beta = \sigma_z \otimes 1$ are gamma matrices in the Dirac representation, $t =2$ is the hopping strength, $m=-1$ is the mass parameter,  and  $a=1$ is the lattice spacing.
\cite{liu2010model,ryu2012disorder,kobayashi2013disordered, kobayashi2014density} This non-interacting model exhibits
 a bulk band gap in the interval $E_F =[-|m|,|m|]$ and a single Dirac cone in the bulk gap. 
To this model  we add uncorrelated white noise  disorder $u(x)$  
chosen randomly from the interval $\left[ -W/2, \, W/2 \right] $, where $W$ is the disorder strength. 
  In the 3-D limit this model's topological phase transition occurs at $W_c \approx 7.5$ when the Fermi energy is at the Dirac point.  \cite{kobayashi2013disordered,Sbierski14}  The disorder is added only on the $d$ layers nearest the TI sample's upper boundary, and also the $d$ layers nearest the lower boundary, as shown in Figure \ref{Schematic}.  Each of these layers has the same disordered strength, and the interior is left clean, so the disorder's spatial profile is a step function.

We will present numerical results about the global density of states  $\rho(E)$
and the  conductance $G(E)$.  
The density of states is  defined as $\rho(E) = \mathrm{Tr}(\delta(E-H))$, where $\delta$ is the matrix version of the Dirac delta function.  We calculate $\rho(E) $ in large $400 \times 400 \times 20$ slabs, with periodic boundary conditions in the slab plane, using  the highly scalable Kernel Polynomial method. 
\cite{weisse2006kernel}
 $\mathrm{Tr}(\delta(E-H))$ is approximated with an expansion  in its Chebyshev moments, the resulting sum over moments is truncated at some maximum number of moments, and this truncation is smoothed  using the Jackson kernel.  We verified  convergence by systematically recalculating our results with different numbers of moments, going as high as $10,000$ moments.   
We found  that the  density of states is self averaging so that $10$ samples were sufficient to obtain high accuracy results.

For the conductance we use the Landauer formula $G = G_0 \;\mathrm{Tr}(\bm t^{\dagger}\bm t)$,
where  $G_0 = e^2/h$ is the conductance quantum, and average over 100 statistical realizations.
$\bm{t}$ is 
the TI's transmission matrix, which we compute using the transfer matrix method \cite{pendry92,Slevin01b,kramer2005random}.  We calculate the conductance at zero temperature so only states at the Fermi level $E_F$ contribute, in contrast to finite temperatures where the Fermi level is smeared across a range of order $k_B T$.  Since the critical disorder strength $W_c$ depends on the Fermi level, at non-zero temperature the conductance will have contributions not only from the topological depinned state, but also from the topological pinned state and the non-topological state.  The size of these contributions, and also of the conductance from thermally activated bulk carriers, can be controlled by reducing the temperature.
We minimize leads effects by using metallic leads.  
Each TI site adjacent to the leads is connected to a perfectly conducting 1-D wire, similarly to  network models. 
We use a slab of height $h=20$, length $l = 40$ between the two leads, and width $w=40$ with periodic boundary conditions along this transverse axis.    
The scattering length is less than $40$ for all disorder strengths greater than $W > 2$, so  finite size effects from the sample width and length are  small. \cite{wu14}
Moreover, because we study disordered boundaries whose maximum depth is $d=5$ layers, the two disordered boundary layers are always separated by at least ten layers of pure non-disordered bulk. Changing the clean bulk's depth from $10$ to $15$ while keeping $d=5$ fixed confirms that the conductance is unchanged when the disorder in the boundary is not too large ($W = 3, 6$ at $E_F = 0, 0.25$), but at larger disorder the conductance increases.  Since the clean bulk's depth is equal to $h - 2d = 20 - 2d$, and our numerical results keep $h = 20$ fixed, our results on the conductance's depth dependence underestimate the behavior of a thick slab.  This does not affect our qualitative conclusions. 

  \begin{figure}[]
\includegraphics[width=8.5cm,clip,angle=0]{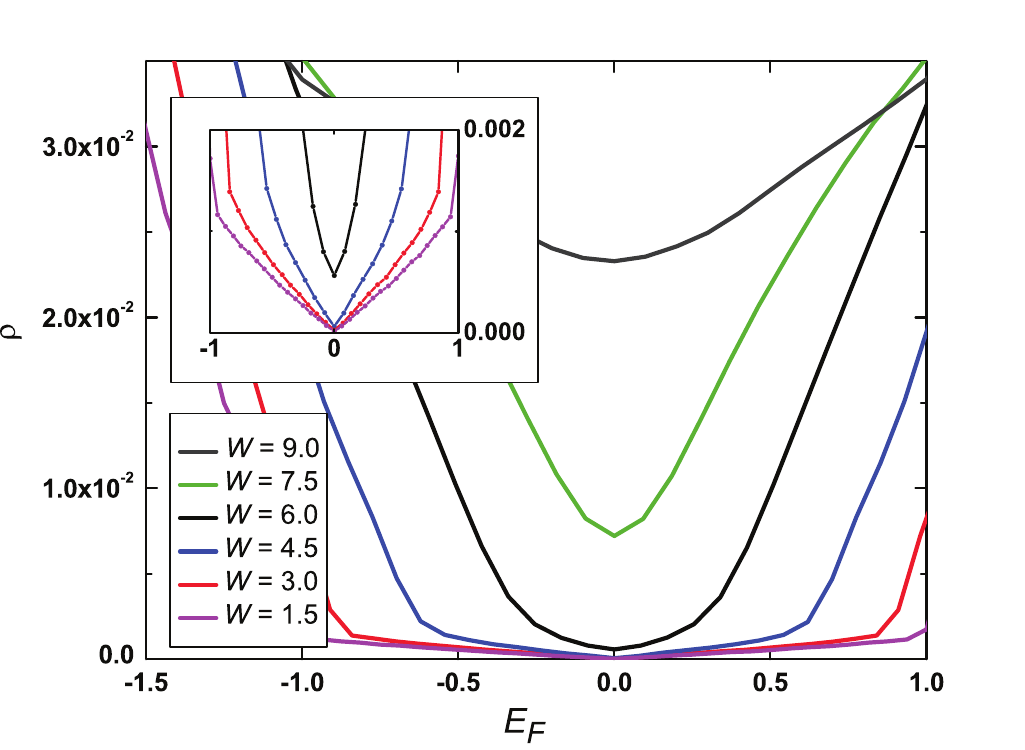}
\caption{ (Color online.)  The normalized density of states inside the bulk gap, evolving from small disorder $W=1.5$ to large disorder $W=9$.  The disorder depth is $d=5$. The inset shows that at  small disorder the DOS is linear in $E_F$, which signals that here the Dirac cone is intact.      }
\label{inGapDOS}
\end{figure}



In Figure \ref{inGapDOS} we focus our attention on the normalized density of states in the bulk gap $E_F = \left[-1,1\right]$, where the topological surface states are found.  
The smallest-disorder  curves are linear in $E_F$, i.e. $\rho\propto |E_F|$, which is a hallmark of  2-D Dirac fermions.
As seen in the inset, the  slope grows with increasing disorder,    because disorder causes a decrease in the Fermi velocity $v_F = dE/dk$.  At larger disorder  $W \geq 3$  the DOS departs from the linear Dirac form in two intervals near $E_F = \pm 1$, and by $W \geq 6$ these intervals   expand to fill the whole band gap.  In these intervals the topological state consecutively becomes strongly disordered, depins, and then is joined by non-topological states in the disordered region.  \cite{Schubert12, wu2013robust}  In particular,  depinned states occur at two critical energies $E_c(W)$ which in the disorder-free $W=0$ case  lie at the band edges $E_c = \pm 1$ but move  inward toward the Dirac point as the disorder grows stronger.  At $W=W_c$ the critical energies meet at the Dirac point  $E_F  = 0$, and at larger $W> W_c$ the disordered region no longer hosts topological states.  The detailed values of $E_c, W_c$ are material dependent, but the qualitative behavior described here is generic to every topological insulator.

\begin{figure}[]
\includegraphics[width=9.5cm,clip,angle=0,trim= 15 10 80 15]{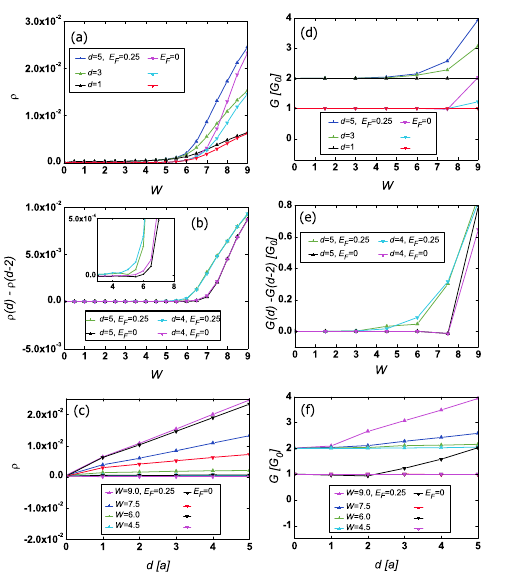}
\caption{ (Color online.)  Transport signatures in the DOS $\rho$ (left panels) and conductance $G$ (right panels)  at two values of the Fermi level $E_F=0,0.25$.   We  have shifted the $E_F=0$ conductance downward by $G_0$ for clarity. 
The pinned topological state is indicated  when $\rho$ is independent of the  disorder depth $d$. Non-topological states are indicated when $G$ is linear in $d$.  The depinned topological state shows a constant $G$ and linear $\rho$.   Panels (a) and (d) show the dependence on $W$.
 Panels (b) and (e) show the change in $\rho$ and $G$ caused by adding two disorder layers. Panels (c) and (f) demonstrate linearity in $d$.}
\label{Conductance}
\end{figure}

Figure \ref{Conductance} shows  signatures of the three types of conducting states in the dependence on depth $d$  of the DOS $\rho$ (left panels) and the conductance $G$ (right panels). 
 Panels (a) and (d) show that at small disorder both $\rho$ and $G$ are independent of $d$, as expected from the $W \ll W_c$ pinned topological state.    At stronger disorder both quantities become sensitive to $d$, as expected of  $W \gg W_c$ non-topological states in the strongly disordered region. Panels (c) and (f) demonstrate that the $d$ dependence is always roughly linear.   
 
 Panels (b) and (e)  allow us to pinpoint the transition  from no $d$ dependence to linear dependence, by plotting  the magnitude of the change when $d$ is changed from $2$ to $4$, and from $3$ to $5$.  Panel (b) shows that at $E_F = 0, 0.25 $ the DOS becomes dependent on $d$  starting at $W \approx 6.5, 5.5$, and the $d$ dependence becomes large around  $W \approx 7.0, 6.5$.   This transition signals  depinning from the TI surface, shown in Figure \ref{Eigenfunctions}b.   
Panel (e) shows that $G$ remains constant in $d$ at $W = 6$ for $E_F = 0$, and nearly constant also at $W=7.5, E_F = 0$.  In this region we have constant $G$ and linear $\rho$, which is the signature of the depinned topological state. 

In summary, we have confirmed numerically that the depinned state's conduction obeys the depth dependence   in Table \ref{table:scaling}, which is based on the fact that this state is both very strongly disordered and robustly conducting.  In consequence its scattering time  $\tau$ scales linearly with the disorder depth $d$ and its scattering length is independent of $d$.  This determines the depth dependence of  all other conduction parameters.  

\section{Applications \label{Applications}} Using this effect, a topological state's conduction  can be engineered and patterned to meet device requirements by introducing a layer of strong $W \approx W_c$ surface disorder, and patterning the layer's depth $d$.  $d$ must substantially exceed the bulk penetration depth $\lambda$, which is typically $2-3 \, \mathrm{nm}$ in the $\mathrm{Bi}_2 \mathrm{Se}_3$ family of TIs. \cite{zhang2010first}   It is necessary that the disorder's spatial distribution has a step function profile; the disorder strength should be constant from the surface up to depth $d$, and then drop to zero.  Ion implantation produces a gaussian depth distribution around a mean depth controlled by the beam energy; a step function distribution may be approximated by applying the beam twice at two different beam energies.

 Increasing the density of states of the in-gap surface states will make them less sensitive to  bulk defects,  which are known to cause  bulk conduction by introducing carriers. Depending on the defect type,  the defects shift the Fermi level toward either the valence or conduction band.   \cite{scanlon2012controlling, cava2013crystal}   When disorder is used to increase the surface density of states the Fermi level will be less sensitive to bulk defects and shift further into the gap, increasing the TI quality.     In patterned devices  points   with increased disorder depth $d$ and density of states $\rho$ will respond more strongly to external static or ac voltages.  Interaction with light also will be increased if the light's penetration depth exceeds $\lambda$. \cite{mciver2012control}
   Moreover the state's self-interaction will be increased, which favors transitions to strongly interacting topological phases. \cite{PhysRevB.83.195139,  PhysRevLett.105.246809}

Engineered lines of increased $d$ on a TI surface can be used to direct the topological current's flow across the  surface, and  to divide current flow and later reunite it, similarly to integrated circuits in conventional semiconductor devices.   These channels can be controlled by using external gates to locally shift the Fermi level.   Because the critical disorder $W_c$ is sensitive to the Fermi level $E_F$, if the Fermi level is either too large or too small then the topological state will reroute to the  boundary of the disordered region and will lose its depth dependence.   Therefore external gates can control the density of states at specific points on the TI surface, regulating the current flow through channels, or switching current from one channel to another.  In summary, the  topological current can be focused, directed along particular channels, and switched, supplying all of the components necessary for realizing topological integrated circuits.

We conclude by discussing a specific device, a spectral analyzer of  incoming  transient pulses.  Its crucial component is a region where the disorder depth $d$ grows continuously.   A topological state diffusing through this region will experience a spatially non-uniform scattering time $\tau$. Its diffusion is similar to Brownian motion of classical particles in the presence of a temperature gradient, since the time between Brownian steps varies inversely with temperature.  As is well known from the celebrated Ludwig-Soret thermodiffusion effect,  the diffusing state will experience an effective force deflecting it along the gradient of $\tau$ and $d$. \cite{platten2006soret, kim2013einstein, rahman2014thermodiffusion}
 The deflection is   greatest when the topological state's energy $E$ is at the critical value $E_c$ associated with depinning, so the spectral analyzer will spatially resolve incoming pulses according to their component energies.

\begin{acknowledgments}
We thank Quansheng Wu, Liang Du, Alex Petrovic, and Ken-Ichiro Imura for very useful discussion and collaboration, and Ivan Shelykh for his support.   This work has been supported by Grants-in-Aid for Scientific Research (B) (Grant No. 15H03700) and Grants-in-Aid No. 24000013.
Part of the numerical calculation has been performed on Supercomputer system B of ISSP, Univ. Tokyo.
\end{acknowledgments}

\bibliography{VincentTO}

\end{document}